# THE REALITY OF CLIMATE CHANGE: EVIDENCE, IMPACTS, AND ENGINEERING SOLUTIONS

Sihua Lu

October 16, 2024

## ABSTRACT

Climate change is one of the most significant global challenges, yet misconceptions persist regarding its causes and impact. This report addresses common myths surrounding climate change and presents scientific evidence to clarify its reality. Utilising data from NASA, NOAA, and the NSW government, this study provides evidence of rising global temperatures, melting ice sheets, rising sea levels, and extreme weather patterns in regions like New South Wales. The analysis demonstrates the human-driven nature of climate change, primarily caused by increased carbon emissions. Engineering solutions, including renewable energy technologies, green buildings, and carbon capture methods, are essential to mitigating the effects of climate change. Future research should focus on improving the scalability of these technologies and addressing the broader impact on ecosystems and human societies.

*Keywords* Climate change · global warming · renewable energy · carbon capture · sea level rise · New South Wales

## 1  Introduction

Nowadays climate change is one of the most pressing world-wide issues. Climate change is a long-term shift in both temperatures and weather patterns. From changing weather patterns that threaten food production, to rising sea levels that increase the risk of catastrophic flooding, the impacts of climate change are global and unprecedented in scale (United Nation, 2024). However, many people remain sceptical about the reality and causes of climate change, though a broad consensus had been agreed in science. For example, some people reckon that global warming is common as a part of natural cycle of climate changing. However, compared to the course of Earth's 4.5-billion-year history, the nowadays rapid warming can't be explained by the cycles of warming and cooling (World Wild Fund, 2024). Some people also think climate change is a future problem, but the extreme weather caused by climate change has authentically decreased the food production, even worse, causing food crisis in past decade years (U.S. Mission Italy, 2024). This paper will devote to explore common misconceptions about climate change and the pseudo-science behind them, as well as provide scientific evidence in support of climate change, and conclude with a discussion of possible solutions to this global crisis.

## 2  Evidence of Climate Change

### 2.1  Global Temperature Rise

One of the most direct pieces of evidence for climate change is the consistent rise in global temperatures. According to NASA's 2024 findings, Earth's average surface temperature has risen by about 1.18°C since the late 19th century, a change driven largely by increased carbon dioxide ($CO_2$) and other human-made emissions into the atmosphere (NASA, 2024). Most of this warming has occurred in the last 40 years, with the seven most recent years being the warmest on record.

There are several methods to measure the temperature rising, including modern devices like thermometers and satellites, also the indirect methods like examining ice cores and tree rings. These technologies provide historical climate data, demonstrating that the current warming trend has occurred repeatedly over millennia. Previous climatic changes were caused by natural causes such as volcanic eruptions and variations in solar radiation, while today's warming is primarily due to human activity, particularly the use of fossil fuels (NASA, 2024).

## 2.2   Melting Ice Sheets, Glaciers and Rising Sea Levels

Melting glaciers and ice sheets is the one most obviously resulting from global warming. Clearly indicating climate change in the modern age is the fast loss of ice in areas like Greenland and Antarctica.
For example, the Greenland ice sheet is shedding ice at concerning speed. Between 1993 and 2019 Greenland lost an average of 279 billion metric tonnes of ice, according to NASA's satellite data, which greatly contributes to sea-level rise (NASA, 2024). With an average of 148 billion metric tonnes lost annually over the same period, Antarctica is likewise losing ice (NASA, 2024).
From the Himalayas to the Andes, glaciers all around are recedes. Millions of people whose water supply depends on glacial meltwater are seriously threatened by the melting of ice blocks. Furthermore, harming coastal towns and ecosystems, glaciers' retreat helps to contribute to increasing sea levels (NOAA, 2023).
Rising sea levels provide still another compelling sign of climate change. The data of U.S. National Oceanic and Atmospheric Administration show that global sea levels have risen by approximately 8-9 inches (21-24 cm) since 1880, and nearly one-third of this rise has occurred in the past twenty-five years. Low-lying island nations like the Maldives and Tuvalu are particularly vulnerable to sea level rise, as invading oceans endanger their basic survival (NOAA, 2024). Furthermore, rising sea levels cause storm surges during typhoons and hurricanes to become more severe, resulting in population displacement, damage of infrastructure and death. Models of climate change project that sea levels will keep rising, with perhaps disastrous consequences should global temperatures fail to stabilise (NASA, 2024).

## 2.3   Ocean acidification

Beyond rising temperatures, oceans are confronting another major problem: growing acidity brought on by higher carbon dioxide ($CO_2$) levels in the atmosphere. About a third of the $CO_2$ emissions from human activities are absorbed by the seas, which sets off a sequence of chemical processes lowering the pH of saltwater (NOAA, 2023.). This process, sometimes referred to as ocean acidification, presents a great threat to marine life particularly coral, molluscs, and some plankton that create their shells or skeletons from calcium carbonate.
Rising acidity makes it more difficult for these species to create and preserve their shells, which raises their mortality rates. Particularly threatened are coral reefs like Great Barrier Reef, sometimes known as the "rainforests of the ocean" because of their great richness. The various marine life that relies on reefs for food and protection suffers as they deteriorate, therefore upsetting more general marine ecosystems (NASA, 2024).

## 3   Climate Change of New South Wales

Not only is climate change obviously and continuously altering weather but also ecosystems, infrastructure, and New South Wales (NSW) general way of life. NSW became more vulnerable because of environmental problems, for example, higher temperatures, shifting rainfall patterns, and bushfires. As a varied landscapes and natural beauty state, NSW is especially vulnerable to climate change, and the extreme climate is changing the state's ecosystem in several ways (NSW Government, 2024).

The higher average temperatures are the most tangible impacts of climate change in NSW. The area has had more regular and severe heat waves over recent years; these are expected to get even more common going forward (NSW Government, 2024). Apart from posing health hazards and discomfort for the population, these increased temperatures have major ecological effects. Long-term heat stress, for instance, can degrade ecosystems, reduce plant and animal populations, and raise the likelihood of drought. NSW's agricultural output is under danger as well as it gets warmer since cattle and crops find it difficult to adapt to the changing temperature (NSW Government, 2024).

Climate change is also causing changes in NSW rainfall trends. Extreme rainfall events and severe drought have both affected the area's local water supply and agricultural systems, therefore upsetting NSW Government, 2024. While excessive rain can cause flooding and soil erosion, hence further damaging infrastructure and the environment, water shortage becomes a serious problem for towns, agriculture, and ecosystems during times of drought (NSW Government, 2024). These erratic weather patterns make planning difficult for farmers and legislators, which adds to the difficulties presented by a changing climate (NSW Government, 2024).

Furthermore, aggravating extreme weather events in NSW is climate change. Heat waves and bushfires, for instance, are becoming common and intense and endanger individuals as well as the natural surroundings. Although Australian terrain has always included bushfires, their growing frequency and intensity in NSW are directly related to climate change (NSW Government, 2024). Especially in woods, increasing temperatures raise the likelihood of major fires. Further aggravating the issue is the expected increasing frequency of extreme weather occurrences including thunderstorms and lightning strikes, which provide natural ignition sources for bushfires as global temperatures rise (NSW Government, 2024).

These weather-related difficulties draw attention to NSW's more general environmental effects of climate change. Disruption of ecosystems all throughout the state causes organisms to struggle to fit new environments. Already experiencing habitat degradation from human activity, native flora and wildlife are under more stress from increasing temperatures and altered precipitation patterns (NSW Government, 2024). Many vulnerable species suffered greatly during the terrible wildfire season of 2019–2020, with some populations driven almost towards extinction. Crucially for preserving the state's natural and agricultural systems, biodiversity loss influences ecosystem services like pollination,
water purification, and soil health (NSW Government, 2024).

Increasingly clear are NSW's social and financial effects from climate change. Extreme weather disasters like bushfires seriously tax people's finances as well as those of the government. Recovering efforts strain public resources and insurance systems and cost billions of dollars, therefore taxing public coffers (NSW Government, 2024). Furthermore, endangering public health is climate change; high heat raises the incidence of heat-related diseases while poor air quality during bushfire seasons aggravates respiratory ailments (NSW Government, 2024).

In order to facing these growing challenges, the NSW government is actively researching and implementing solution to adapt to climate change. Through initiatives such as improved fire management, water conservation, and climate-resilient infrastructure planning, policymakers are working to mitigate the effects of a warming climate on the state's population and environment (NSW Government, 2024). The government is also investing in renewable energy and sustainability programs to reduce greenhouse gas emissions and slow the progression of climate change, aiming to protect future generations from its worst impacts (NSW Government, 2024).

## 4 Engineering Solutions to Fight Climate Change

### 4.1 Renewable Energy Technologies

Probably the most relevant and realistic scientific and engineering contribution that could be thrown in to fight climate change is in the form of renewable energy technologies. Renewable energy includes solar, wind, geothermal, and tidal power, to name a few. These are essentially necessary as it enters the greenhouse gases reduction targets; two-thirds of global emissions emanate from the production of energy. Engineering advancements in these areas are making these renewable energy sources more efficient and affordable, enabling quicker deployment on a larger scale. Over the past decade, solar and wind energy have seen significant cost reductions, making them the most cost-effective forms of power generation in many regions around the world (Mannan, 2021).

However, the current rate of renewable energy deployment must increase sixfold to meet the climate mitigation targets set out in the Paris Agreement [2]. Engineers are continuously working on enhancing grid management systems and expanding energy storage technologies, like batteries, to accommodate the variability of renewable energy sources [2]. Solar and wind farms can be built in less than a year, compared to the five years or more required for fossil fuel plants, making renewable energy highly scalable. Furthermore, these energy systems can be deployed in various environments, including offshore or in remote areas, offering a flexible solution for both developed and developing nations (Mannan, 2021).

### 4.2 development of cleaner transportation systems

Hybrid and pure electric vehicles have only just started cleaning up transport, but a lot of work is still to be done in transitioning the infrastructure and making EVs more usable. Further, innovations in smart grid management and better urban planning can help further reduce emissions by supporting the adoption of electric vehicles and public transportation systems.

In the case of aviation, the engineering challenge is to find alternative fuels or technologies that can make flying carbon neutral. One feasible solution is hydrogen fuel cells for aircraft. Airbus, one of the world's largest makers of airplanes, is currently working on the first zero-emission aircraft; it plans to launch no earlier than 2035. While decarbonization in aviation is quite a difficult process, ongoing research and innovation mean that cleaner aviation technologies could feasibly be widely deployed over the coming decades.

### 4.3 Green buildings

Green buildings also play a crucial role in combating climate change. Buildings are significant contributors to greenhouse gas emissions, both during construction and throughout their operation. Engineering solutions for green buildings focus on minimizing energy consumption and maximizing the use of sustainable materials. The World Green Building Council defines green buildings as those that reduce or eliminate negative environmental impacts through efficient use of energy, water, and materials [2].Engineers are developing structures with natural ventilation, renewable energy heating and cooling systems, and either recyclable or sustainably sourced materials.

To evaluate building sustainability, several rating systems—including LEED (Leadership in Energy and Environmental Design) have been created (Mannan, 2021). These technologies inspire builders and designers to use greener methods by measuring elements including energy efficiency, trash reduction, and air quality. Over time, green buildings can lower the total carbon footprint of metropolitan areas and raise resident quality of living simultaneously.

### 4.4 Direct Air Capture Carbon Capture

Developing direct air carbon capture technologies—which seek to directly collect carbon dioxide (CO2) from the atmosphere—perhaps represents one of the most aspirational engineering solutions. Although still in its early years, this technology has showed promise as a possible means of attaining negative emissions—that is, where more CO2 is taken out of the environment than is released [1]. By sending air through a system designed to remove CO2 and either bury it underground or use it for industrial purposes, carbon capture devices operate Although present carbon capture technologies are costly and energy-intensive, continuous research aims to make the process more scalable and reasonably priced [1].

For example, Harvard engineer Aaron Sabin is creating a new method for carbon capture that separates CO2 from the air using electrochemical techniques. Sabin is a potential large-scale deployment [1] solution since her approach uses more energy than conventional heat-based carbon capture technologies. As such technologies advance, they may become rather important in lowering atmospheric CO2 levels and enabling the reversal of climate change.

## 5 Conclusion

Dealing with climate change calls for a multifarious strategy combining policy-driven solutions, scientific, technical, and engineering aspects. Although solar and wind power among other renewable energy sources is quite important in lowering greenhouse gas emissions, their quick implementation must be given top priority if we are to reach world climate targets. Likewise, developments in hybrid and electric cars as well as in transportation infrastructure have great possibilities to reduce emissions, especially in the very polluting aviation industry. Sustainable urban design and green buildings help to reduce our environmental effect even further, therefore strengthening the resilience and energy economy of cities. Concurrent with this, the evolution of technology such as direct air carbon capture offers a window into the future of carbon elimination, so affording hope that we can not only slow but also reverse the negative consequences of climate change. Though obstacles still exist, our efforts to build a sustainable future depend on ongoing innovation and cooperation among engineers, scientists, and legislators. Working together to use these ideas will help us to significantly solve one of the most important worldwide issues of our day.

## References

[1] Mannan, R. (2021). Engineering solutions to fight climate change. Retrieved October 3, 2024, from https://newengineer.com/blog/engineering-solutions-to-fight-climate-change-1513622

[2] Massari, P. (2023). Engineering a solution to climate change. Retrieved October 3, 2024, from https://gsas.harvard.edu/news/engineering-solution-climate-change

[3] NASA. (2024). Evidence. Retrieved October 1, 2024, from https://science.nasa.gov/climate-change/evidence/

[4] NOAA. (2024). Understanding ocean acidification. Retrieved October 1, 2024, from https://www.fisheries.noaa.gov/insight/understanding-ocean-acidification

[5] NSW Government | AdaptNSW. (2024). Climate change impact on bushfires. Retrieved October 3, 2024, from https://www.climatechange.environment.nsw.gov.au/bushfires

[6] United Nations. (2024). Climate change. Retrieved October 1, 2024, from https://www.un.org/en/global-issues/climate-change

[7] U.S. Mission Italy. (2024). How climate change affects the food crisis. Retrieved October 1, 2024, from https://it.usembassy.gov/how-climate-change-affects-the-food-crisis/

[8] World Wildlife Fund. (2024). 10 myths about climate change. Retrieved October 1, 2024, from https://www.wwf.org.uk/updates/here-are-10-myths-about-climate-change